\documentclass[conference,a4paper]{IEEEtran}
\IEEEoverridecommandlockouts

\usepackage[ruled,linesnumbered]{algorithm2e}

\def\BibTeX{{\rm B\kern-.05em{\sc i\kern-.025em b}\kern-.08em
    T\kern-.1667em\lower.7ex\hbox{E}\kern-.125emX}}
    
\usepackage{CJKutf8}  

\usepackage{xcolor,colortbl}
\usepackage{pgfplots}
\usepackage{tikz}
\usepackage{comment}

\usepackage{footnote}
\makesavenoteenv{tabular}
\makesavenoteenv{table}

\usepackage{filecontents}

 \usepackage{threeparttable}
\usepackage{fancyhdr} 
\usepackage{hyperref}
\hypersetup{
    colorlinks=true,
    linkcolor=blue,
    filecolor=magenta,      
    urlcolor=black,
    pdftitle={An Example},
    pdfpagemode=FullScreen,
    }
\usepackage{multirow}
\usepackage{cite}
\usepackage{amsmath,amssymb,amsfonts}

\usepackage{textcomp}

\usepackage{subfigure}
\usepackage{pgfplots}
\usepackage{filecontents}

\usepackage{graphicx,subfigure}

\usepackage{amsthm}
\theoremstyle{plain}

\newtheorem{defi}{Definition}

\definecolor{bblue}{HTML}{4F81BD}
\definecolor{rred}{HTML}{C0504D}
\definecolor{ggreen}{HTML}{9BBB59}
\definecolor{ppurple}{HTML}{9F4C7C}

\usepackage{listings}

\usepackage{blindtext}
\usepackage{etoolbox,xstring,mfirstuc,textcase}
\usepackage{booktabs}
\usepackage{pgfplots}

\usepackage{tabularx,booktabs,makecell,multirow, caption}
\usepackage{enumitem} 

\newcolumntype{L}{>{\arraybackslash}X}

\usepackage{xcolor}
\usepackage{tikz}
\usepackage{pgfplots}

\definecolor{findOptimalPartition}{HTML}{D7191C}
\definecolor{storeClusterComponent}{HTML}{FDAE61}
\definecolor{dbscan}{HTML}{ABDDA4}
\definecolor{constructCluster}{HTML}{2B83BA}

\usepackage{listings}
\begin{document}
\title{Rational Ponzi Game in Algorithmic Stablecoin}

\author{\IEEEauthorblockN{Shange Fu\IEEEauthorrefmark{1}, Qin Wang\IEEEauthorrefmark{2}, Jiangshan Yu\IEEEauthorrefmark{1}, Shiping Chen\IEEEauthorrefmark{2}
}

\IEEEauthorrefmark{1}  \textit{Monash University, Australia}\\
\IEEEauthorrefmark{2} \textit{CSIRO Data61, Australia}
}

\maketitle

\begin{abstract}
Algorithmic stablecoins (AS) are one special type of stablecoins that are not backed by any asset.
They stand to revolutionize the way a sovereign fiat operates.
As implemented, AS are poorly stabilized in most cases; their prices easily deviating from the target or even falling into a catastrophic collapse, and are as a result often dismissed as a Ponzi scheme.
However, what is the essence of Ponzi?
In this paper, we try to clarify such a deceptive concept and reveal how AS work from a higher level.
We find that Ponzi is basically a financial protocol that pays existing investors with funds collected from new ones.
Running a Ponzi, however, does not necessarily imply that any participant is in any sense losing out, as long as the game can be perpetually rolled over.
Economists call such realization as a \textit{rational Ponzi game}.
We thereby propose a rational model in the context of AS and draw its holding conditions.
We apply the model to examine: \textit{whether or not the algorithmic stablecoin is a rational Ponzi game.}
Accordingly, we discuss two types of algorithmic stablecoins (\text{Rebase} \& \text{Seigniorage Shares}) and dig into the historical market performance of a number of impactful projects to demonstrate the effectiveness of our model.

\end{abstract}


\section{Introduction}
\label{sec:intro}

Ponzi is making a comeback that is used to describe the hyped algorithmic stablecoins today. The most recent critique was directed at the one called TerraUSD (UST), which crashed by almost 98\% within 24 hours on May 12$^{\text{th}}$, 2022 (cf.\cite{coingecko}), impacting a direct of \$85B market value evaporation and the collapse of entire crypto markets. This worst-hit reveals a new type of bank run and has triggered repetitive rounds of criticism and discussion about the stablecoin and its algorithmic stabilized approaches. Stablecoins are created to inherit all great features that cryptocurrencies hold (e.g., decentralization, automation, cross-border), without, however, suffering from the same volatility, making them much more applicable as a store of value, medium of exchange, and unit of account~\cite{sokemin,stablecoins2.0,demys_stable}. In an ideal form, stablecoins are simply cryptocurrencies with stable value. Current stablecoin implementations (Tab.\ref{tab:stablecoin}) aim to bridge external assets with the cryptocurrency space by anchoring a peg (e.g., USD) in price. 

Stablecoins are generally tagged as either collateralized or algorithmic stablecoins based on whether they are backed by collateral or not.
The collateralized stablecoin, as the name indicates, requires collateral and thus ensures the circulating token has a redemption value.
Major collateral options include fiat money, cryptocurrency, and other valuable assets.
Currently, stablecoins are primarily backed by fiat money (Tab.\ref{tab:stablecoin}).
USDC \cite{usdc}, for example, is a fully collateralized stablecoin pegged to USD.
The USDC issuer claims that they store an equivalent amount of US dollars of the supply of USDC in a list of banks.
It is reasonable to argue that stablecoins backed by fiat money are of a high degree of centralization in essence.
Using cryptocurrency as collateral can instead circumvent this problem. However, considering its high volatility, an \textit{over}-collateralization is often required if relying on cryptocurrency as collateral. One representative in this category is stablecoin DAI \cite{dai}, where a user needs to deposit Ether (the native token of Ethereum blockchain) in excess of the amount of created DAIs.
This way, even if the underlying crypto depreciates, each unit of the stablecoin still has enough space to be redeemed for at least the same value of its peg.

Algorithmic stablecoin, however, does NOT require any collateral.
Its money supply is elastic and managed by algorithms (i.e., coded in smart contracts), and thus not limited in scale.
As for stabilization, the par value of an algorithmic stablecoin is basically preserved by expanding supply when the price is too high and contracting supply when the price is too low.
There are two major designs in algorithmic stablecoins~\cite{volatility}, namely, \textit{Rebase} and \textit{Seigniorage Shares}.
Rebase~\cite{hayekrebase} is an algorithm that automatically adjusts the money supply in response to the market demand on a routine basis.
A rebase protocol can add ($\mathsf{mint}$) or remove ($\mathsf{burn}$) coins from circulation according to the stablecoin's price deviation from the peg, and this is usually achieved by managing token balance pro rata across all wallets.
Instead of rebasing one coin, seigniorage shares~\cite{noteonshares} normally has a \textit{dual} (or \textit{multiple}) coin design: the price-stable coins and flexible investment shares.
The shares can algorithmically adjust the supply of price-stable coins much like a central bank does with fiat currencies.
Seigniorage shares model has, to varying degrees, served as a foundation for TerraUSD \cite{TerraUSD}, Basis Cash \cite{basiscash}, and Frax \cite{frax}.

\begin{table}[!hbt]
\renewcommand\arraystretch{1.2}
\caption{An Overview of Stablecoins (Ranked by Mrk Cap)}
\vspace{-0.1cm}
\label{tab:stablecoin}
\resizebox{1\linewidth}{!}{
\begin{tabular}{l|l l l l r r}
\toprule
\textit{\textbf{Name}} &  \textit{\textbf{Token}} & \textit{\textbf{Centralised}}  & \textit{\textbf{Supply}} & \textit{\textbf{Peg}} & \textit{\textbf{Mrk Cap}}\\
\midrule

 \textbf{Tether}   &  USDT   & Centralised   &  Collateral   &  USD  &  \$157.4B    \\
 \textbf{USD Coin}   &  USDC   & Centralised   &  Collateral   &  USD  &  \$68.5B    \\
 \textbf{BinanceUSD}    & BUSD  & Centralised   &  Collateral   &  USD  &  \$21.5B    \\
 \textbf{Dai}   & DAI  & Decentralised  &  Collateral   &  USD  &  \$6.2B    \\
 \textbf{Frax}   & FRAX   &  Decentralised   &  Fractional Alg.  &  USD  &  \$1.4B    \\
 \cellcolor{pink} \textbf{TerraUSD}   & UST-LUNA  &  Decentralised    &  Alg. (S.Share) &  USD  &  \$471.1M   \\
 \textbf{Olympus}  & - & Decentralised  &  Collateral  &  -  &  \$303.4M    \\ 
 \textbf{Fei USD}   & FEI  &  Decentralised   &  Collateral B.C.   &  USD  &  \$57.2M    \\
 \cellcolor{pink} \textbf{Ampleforth}   &  AMPL  & Decentralised  &  Alg. (Rebase)  &  CPI USD  &  \$45.6M    \\
 \textbf{Empty Set}   &  ESD   &  Decentralised   &  Alg. (S.Share)   &  USD  &  \$370.5K   \\
  \cellcolor{pink} \textbf{Basis Cash}  &  BAC-BAB-BAS   & Decentralised   &  Alg. (S.Share)   &  USD  &  \$281.7K   \\

\bottomrule
\end{tabular}
}
\begin{tablenotes}
       \footnotesize
       \item[]
        \textbf{Data:} Fetched on $12^{\text{th}}$ Dec 2022. \textbf{Abbreviation:} Alg. = Algorithmic; 
       \item[]
        Mrk = Market; S.Share = Seigniorage Shares; B.C. = Bonding Curve\footnote{A bonding curve is a mathematical concept used to describe the relationship between the supply of a token and its price. In its simplest form, the more token that has been issued, the higher the token price is.}.
     \end{tablenotes}
\vspace{-0.2cm}
\end{table}

\smallskip
\noindent\textbf{Contributions.}
However, there is no satisfactory algorithmic stablecoin that has been healthily operated as of now.
Non-collateral property and non-stable performance undermine people's already fragile trust,
and algorithmic stablecoins are thus often questioned as Ponzi schemes.
This motivates our observations towards this fancy but fuzzy concept of Ponzi.
First, we review existing in-the-wild algorithmic stablecoin projects and 
explore the design primitives they share in common (Sec.\ref{sec:intro}).
We then accordingly establish a rational Ponzi model in the context of algorithmic stablecoins (Sec.\ref{sec:model}) and evaluate three representative projects, namely, Ampleforth (rebase), TerraUSD (seigniorage share), and Basis Cash (seigniorage share), based on six-month historical data from Mar 1$^\text{st}$ to Oct 17$^\text{th}$ 2022 (Sec.\ref{sec:apply}).
We also find several interesting insights (Sec.\ref{sec:discuss}) that are driven by our model.
At last, we deliver two taking-home messages saying:

\begin{itemize}
    \item[$\diamond$] \textit{It is more likely to run a rational Ponzi game in decentralized finance (DeFi) protocols than in traditional finance.} In traditional finance, default (a failure to make repayments on a debt) is an intrinsically intractable problem as human being's financial behavior is non-controllable. DeFi, however, empowered by the automatically operated smart contracts, can run functions such as enforceable payment to ensure a Ponzi-style debt to roll over.

     \item[$\diamond$] \textit{Algorithmic stablecoin has a chance to achieve the defined rational Ponzi model under certain conditions.} Based on our investigations, the rebase approach is relatively easy to implement as a rational Ponzi game as long as there always exists an active trading market for the token always exitsts. Whereas the seigniorage share method is of low probability to realize so. This is due to its inability to cope with the joint fall effect caused by the multi-coin design when a death spiral is triggered.
     
\end{itemize}

\noindent\textbf{Recent Studies.} A series of papers and reports~\cite{demys_stable,designstablecoins,stablecoinsurvey,multicoinsry} establish and extend the taxonomies in stablecoins based on the peg type, the collateral type, and the collateral amount. They give an initial sight to demystify its assemblies and components. Besides, with in-depth analysis, a number of follow-up works has been proposed. Moin et al. \cite{sokemin} provide a systematical classification of existing stablecoins in terms of their design elements. Ariah et al. \cite{stablecoins2.0} investigate the economic structure of stablecoins and establish a risk-based functional characterization. Pernice et al. \cite{openquestion} research stablecoins from a monetary perspective. Salehi et al. \cite{dai_liquidation} explore the liquidation process in stablecoins, and put forward a research agenda for alternatives to liquidation. Clements R. \cite{clements2021builttofail} studies the inherent fragility in algorithmic stablecoins and claims that they are built to fail. Zhao et al. \cite{volatility} conduct an empirical analysis to explain the volatility of algorithmic stablecoins.

\section{Modeling Rational Ponzi Game}
\label{sec:model}

Ponzi~\cite{ponzi_wiki} has always been such a term that is instinctively resented by the public. For malicious organizers, Ponzi schemes are used to lure new investors, and leverage incoming money to pay earlier ones while keeping the rest as profit.
In general, Ponzi organizers normally will promise an abnormally high return to investors within the short term. With little or no legitimate earnings, Ponzi games require a constant cash flow of new funds to survive. Once becoming difficult to recruit new participants, or whenever a large number of existing investors cash out, the game would tend to collapse. 
However, stemming from its origin, the substance of Ponzi is neutral yet subtle. Economists~\cite{ponzi_minsky,ponzi_kindleberger} discuss a perfect foresight version of the Ponzi model, where games based on irrational lenders or imperfect information are ruled out. They call one possibility, in which all debts can be forever \textit{rolled over}, i.e., financed by issuing new debt, a \textit{rational Ponzi game}~\cite{ponzi_rational}.
Following such narrative connotations, we develop a rational Ponzi model in the context of algorithmic stablecoins.

\smallskip
\noindent\textbf{Rational Ponzi Game.}
An entity that plans to issue debt and roll it over the possibly infinite future is the borrowing side of a rational Ponzi game. Any plan will impose a set of net cash inflows $I_s$ from the lending side. The present value of borrower's net indebtedness $D_{T}$ of at any time $T$ equals the present value of the stream of prospective incomes generated by lenders between $0$ and $T$:

\begin{equation}
\Gamma(T) D_{T}=\sum_{s=1}^{T} \Gamma(s) I_{s}
\end{equation}

\noindent where $\Gamma(T)$ is the discount factor\footnote{In corporate finance, a discount factor is a number that is derived from the discount rate. It is used to discount future cash flows back to their present value at a specific discount rate. Here, we assume $r_{t}$ are identical for all assets under a perfect foresight setting.} of the borrowing side, it is calculated by the interest rate $r_{t}$ between periods $t-1$ and $t$, and $\Gamma(s)$ is the discount factor applicable in period 0 to cash inflows received in the period $s$, where $ \Gamma(s) \equiv \prod_{j=1}^{s}\left(1+r_{j}\right)^{-1}.$
We now provide the definition of the rational Ponzi game.

\begin{defi}[Rational Ponzi Game]\label{defi-ponzi}
A rational Ponzi game is a sequence of debt transactions satisfying the following sense: \begin{itemize}
    \item[(i)] to the borrower, it is a positive net present indebtedness value such that $\lim _{T \rightarrow \infty}$ $\Gamma(T) D_T>0$, if the limit of $\Gamma(T) D_T$ exists;
    \item[(ii)] to all the participants, i.e., both the borrowing and lending sides, the game enables a Pareto improvement\footnote{A Pareto improvement is an improvement to a given situation, where a change in allocation of goods harms no one and benefits at least one individual.}.
\end{itemize}
\end{defi}

\begin{figure}[!hbt]
	\centering
	\includegraphics[width=\linewidth]{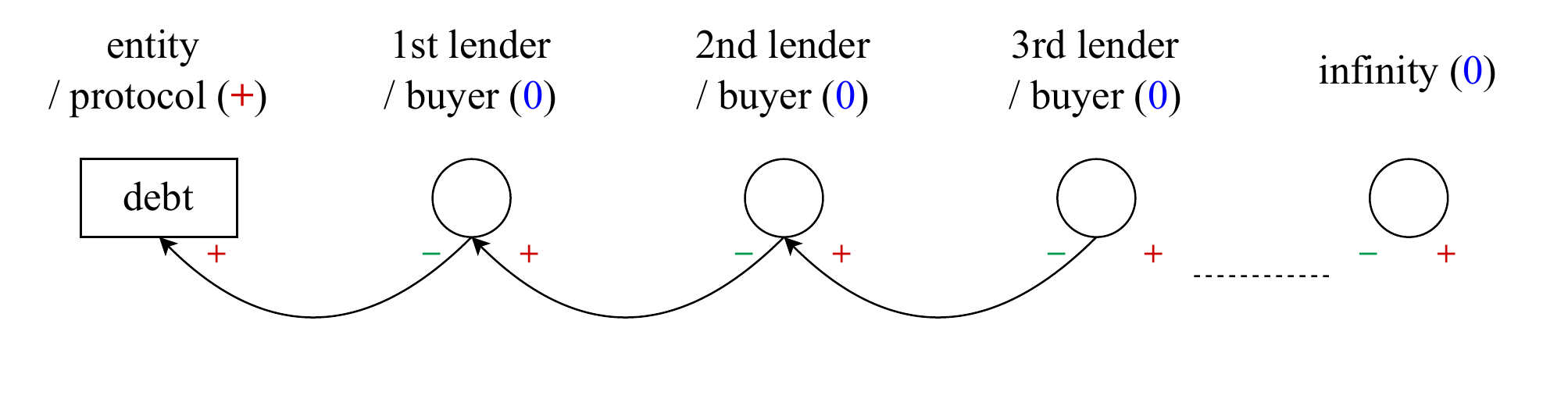}
	\caption{How a rational Ponzi game works.} 
	\label{fig:rational}
\end{figure}

The first holding condition in Definition 1 implies that debts are unnecessarily fully paid off by future cash inflows if in an infinite horizon setting. The second condition imposes restrictions that no one is being worse (debt can always be redeemed at least equal to the market value of assets being deposited). Therefore, the two conditions guarantee a rational realization. We provide a general operation process of the rational Ponzi model as in Fig.\ref{fig:rational}. The red plus sign and the green minus sign represent cash inflow and cash outflow, respectively. The blue values in parentheses following each participant represent their corresponding utility.


\noindent\textbf{Scope of application.} One may consider that only governments are entitled to run such rational Ponzi games. However, in a world where rational Ponzi games can exist, any infinitely lived entity can issue debt and perpetually roll it over. In principle, even a finitely lived entity can issue bonds with zero coupons, i.e., fiat currency. All required is to make participants believe that there will always exist followers being willing to purchase the debt.
At the same time, economists have also proven that the introduction of perpetually rolled-over debt will never make the lending market worse off relative to an economy in which no Ponzi game runs~\cite{ponzi_rational,tirole1985asset,pernice2019monetary}.

\section{Applied to Algorithmic Stablecoin}
\label{sec:apply}

Typically, algorithmic stablecoins, based on their operating mechanisms, can be considered a Ponzi game: an on-chain protocol launches a coin (equiv. debt) claimed to be stable (investors can always get the equivalent fund back when selling the coin), and one joins the game by purchasing the coin and can quit the game if others are willing to buy his coins on the market. However, is such an algorithmic stablecoin game can being rational under our model? In this section, we investigate three representative instances, Ampleforth (rebase), TerraUSD (seigniorage shares), and Basis Cash (seigniorage shares), to answer this question.
Supposing that all the lending protocols are deployed on smart contracts on a public ledger so that all steps are perfectly enforceable (equiv. no default risk). We define the notation $\mathcal{U}$ as the investor's utility; it is basically the product of the number of held assets $Q$ and the market price of the token $P$, in symbols as
\begin{equation}
\mathcal{U} = Q \times P.
\end{equation}

Then, we apply our model to three concrete instances by investigating their six-month historical data (Mar\,--\,Oct, 2022). We analyze the utility of investors based on extracted data covering price, quantity, and market cap.

\smallskip
\noindent\textbf{Ampleforth (Rebase).} Rebasing is a process applied by several algorithmic stablecoins to maintain their peg to a certain value, such as the US dollar.
When a stablecoin rebases, it adjusts the supply of tokens in circulation in response to changes in the value of the underlying asset or assets.
It's important to note that rebasing can be a complex process, and different algorithmic stablecoins may use different algorithms to determine when and how to rebase.
Ampleforth (AMPL) \cite{ampleforth} is one representative protocol that applies the design of rebase.
AMPL is pegged to the CPI-adjusted USD, and its rebase procedure occurs at 02:00 UTC on a daily basis.
The quantity of AMPL tokens in user accounts automatically expands or contracts based on everyday price.
According to Ampleforth whitepaper \cite{amplwhiterpaper}, the absolute \textit{supplyDelta} of AMPL is equal to $\frac{(\mathsf{price}-\mathsf{target})*\mathsf{totalSupply}}{\mathsf{target}}$.
For example, at the time of rebase, if the price of AMPL is \$1.1, 10\% greater than the standard 1 USD peg, a wallet with a balance of 100 AMPL will be rebased as 110 accordingly. In contrast, when the AMPL price is less than 1 USD, the protocol will proportionally decrease the quantity of tokens in wallets, aiming to bring the price up to its anchor.

\begin{figure*}[!hbt]
\subfigure[AMPL price]{\label{fig:ampl-1}
\resizebox{0.33\textwidth}{!}{
	\includegraphics[width=1\textwidth]{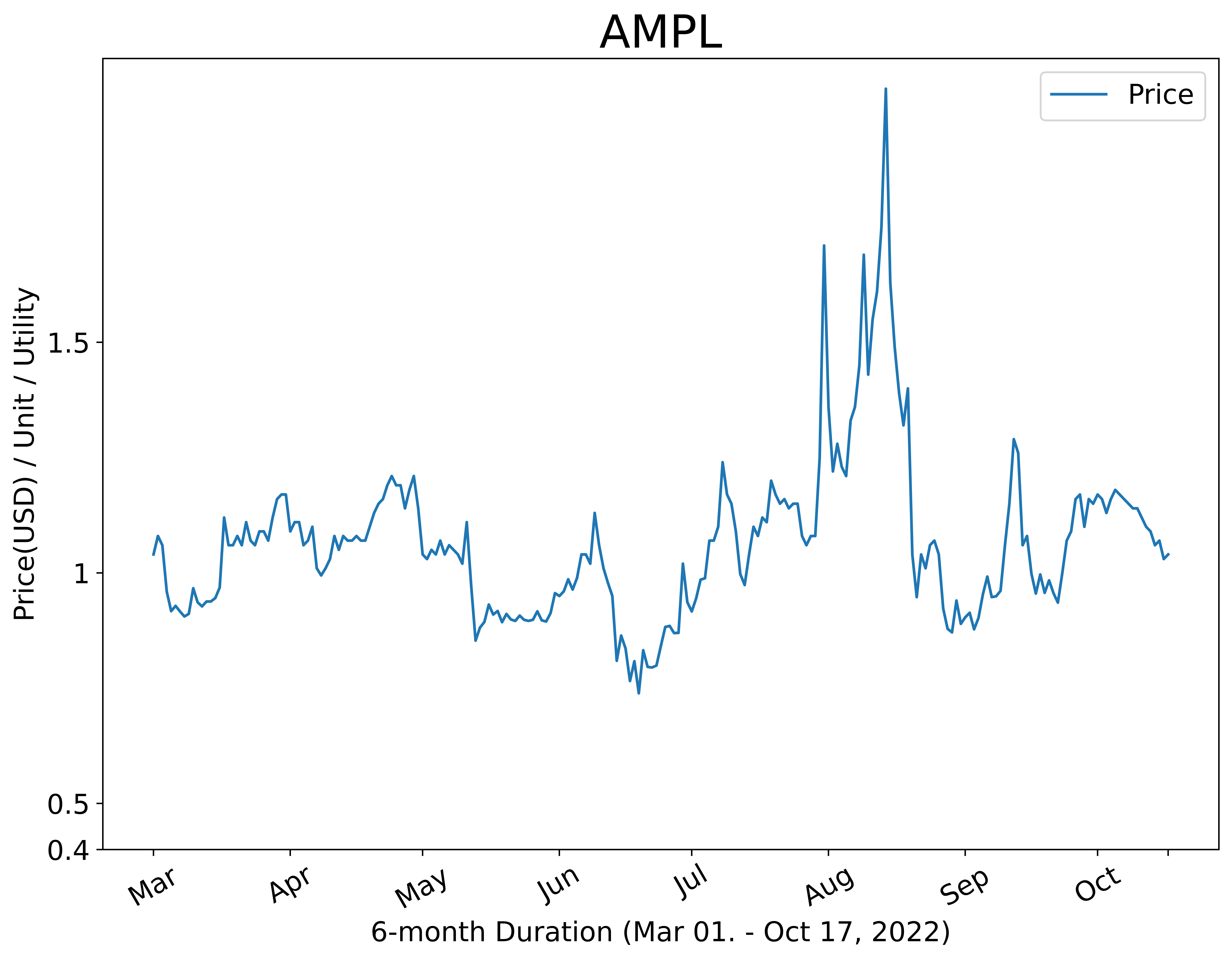}
	}
	}
\subfigure[AMPL amounts]{\label{fig:ampl-2}
\resizebox{0.33\textwidth}{!}{
	\includegraphics[width=1\textwidth]{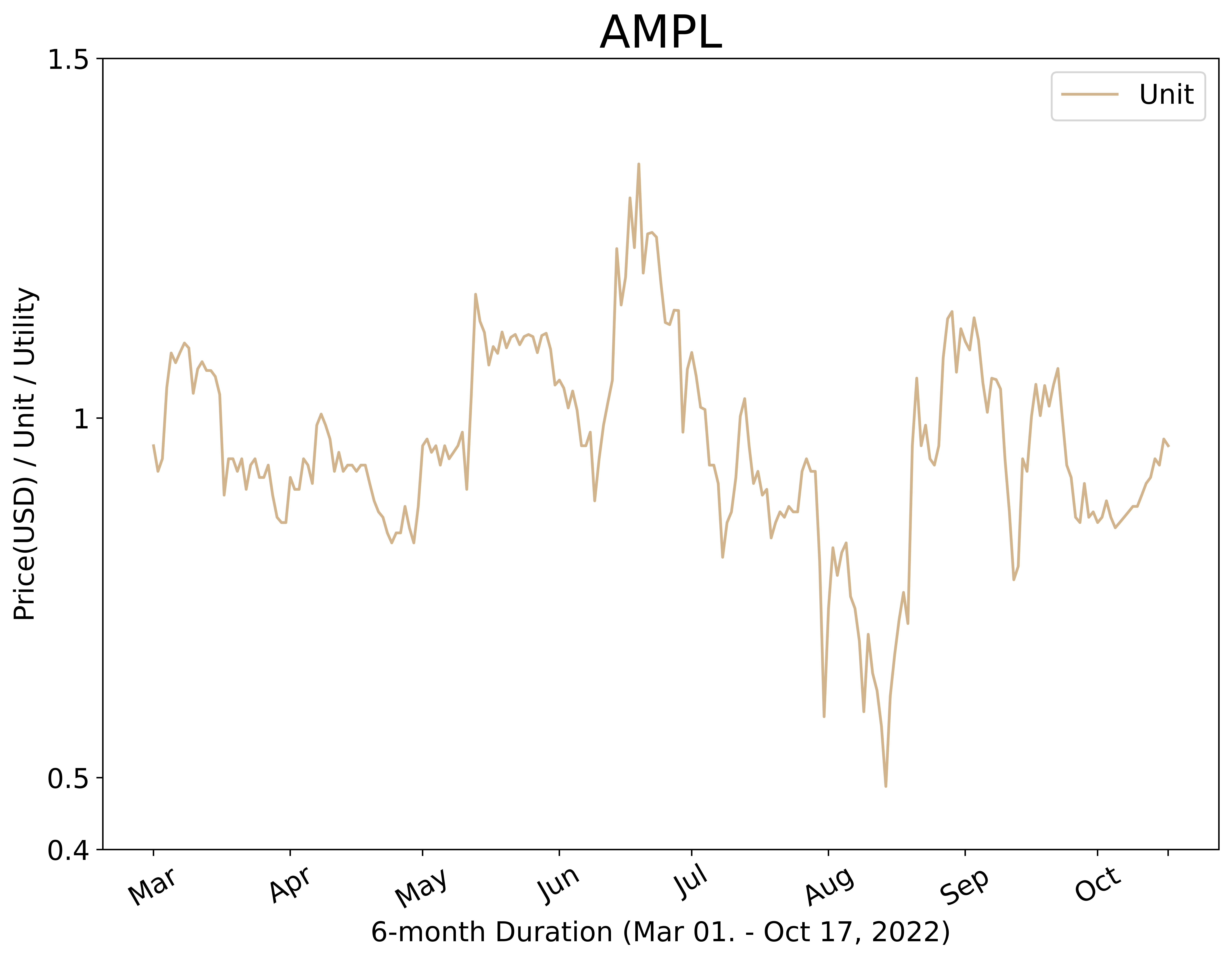}
	}
	}
\subfigure[Investor's utility]{\label{fig:ampl-3}
\resizebox{0.33\textwidth}{!}{
	\includegraphics[width=1\textwidth]{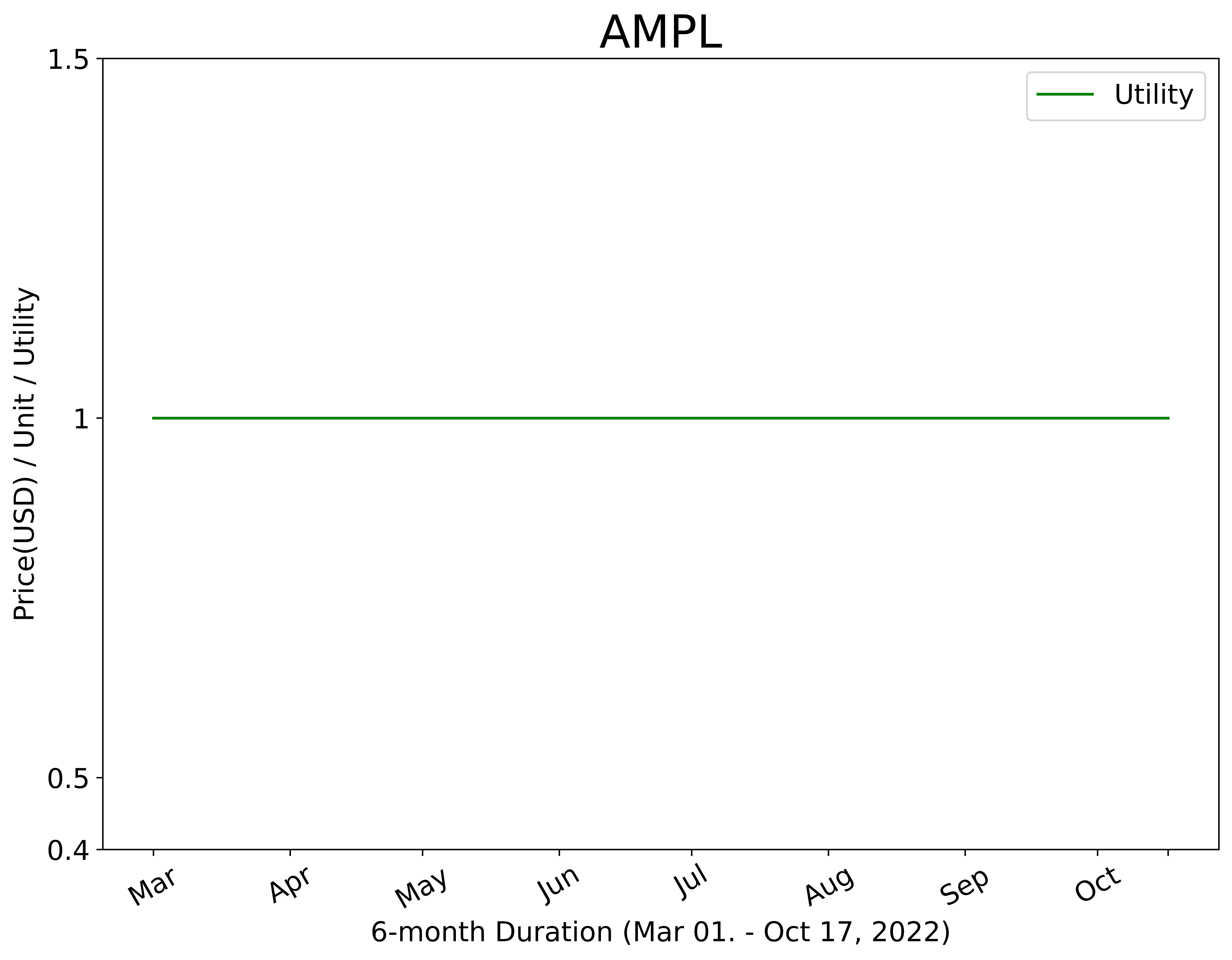}
	}
	}
	\caption{AMPL status over time.}
	\label{fig:ampl}

\subfigure[UST price]{\label{fig:ust1}
\resizebox{0.33\textwidth}{!}{
	\includegraphics[width=1\textwidth]{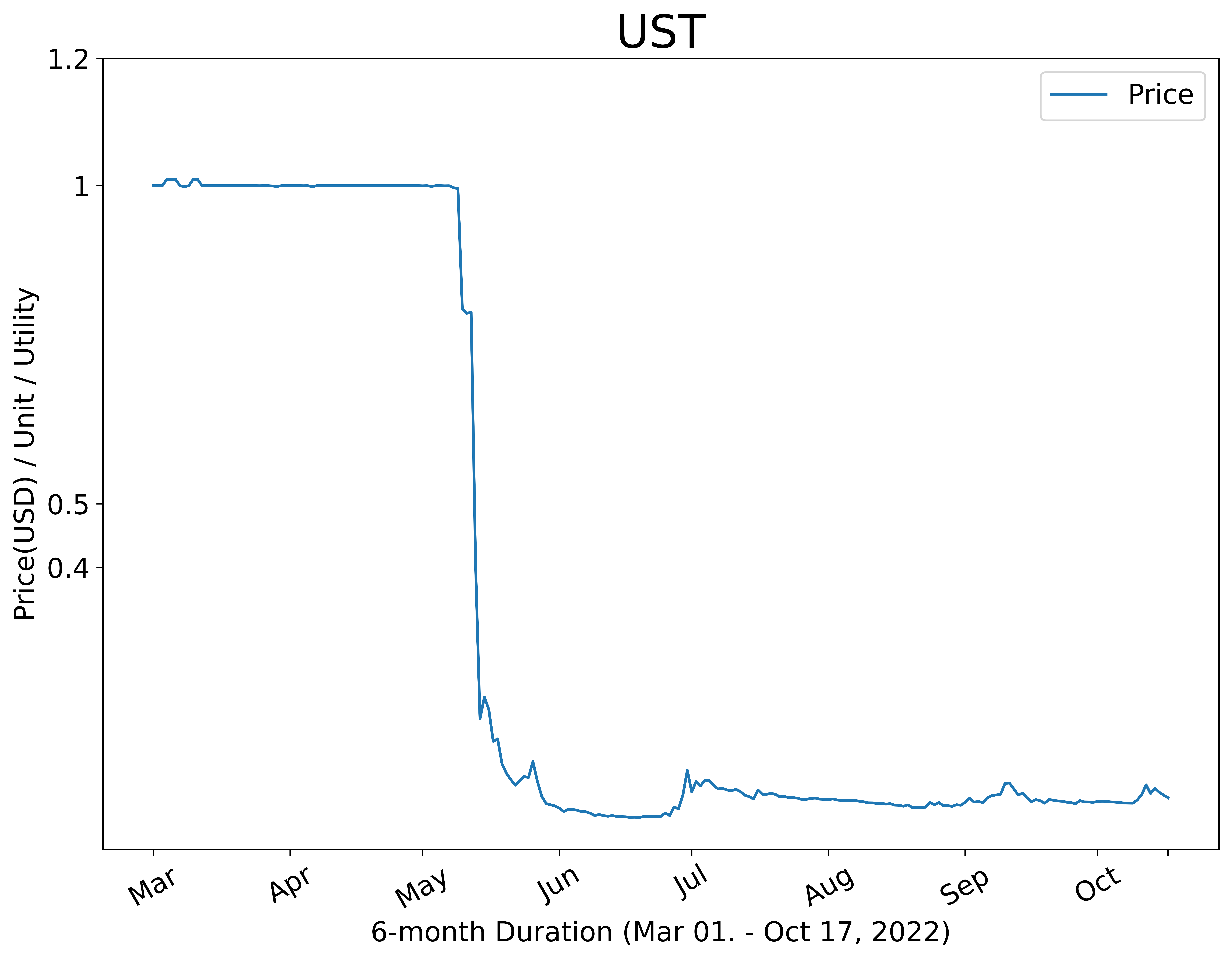}
	}
	}
\subfigure[UST amounts]{\label{fig:ust2}
\resizebox{0.33\textwidth}{!}{
	\includegraphics[width=1\textwidth]{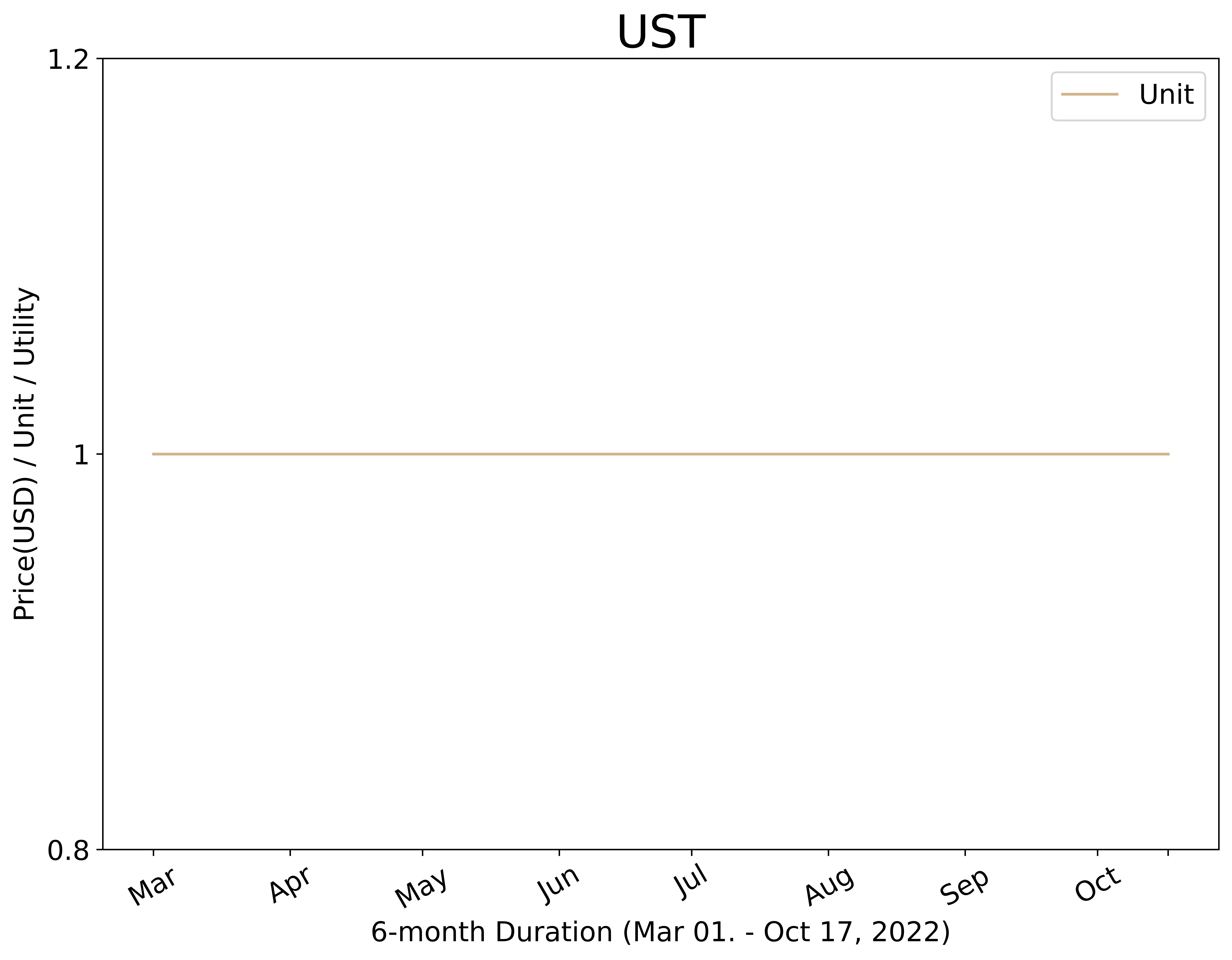}
	}
	}
\subfigure[Investor's utility]{	\label{fig:ust3}
\resizebox{0.33\textwidth}{!}{
	\includegraphics[width=1\textwidth]{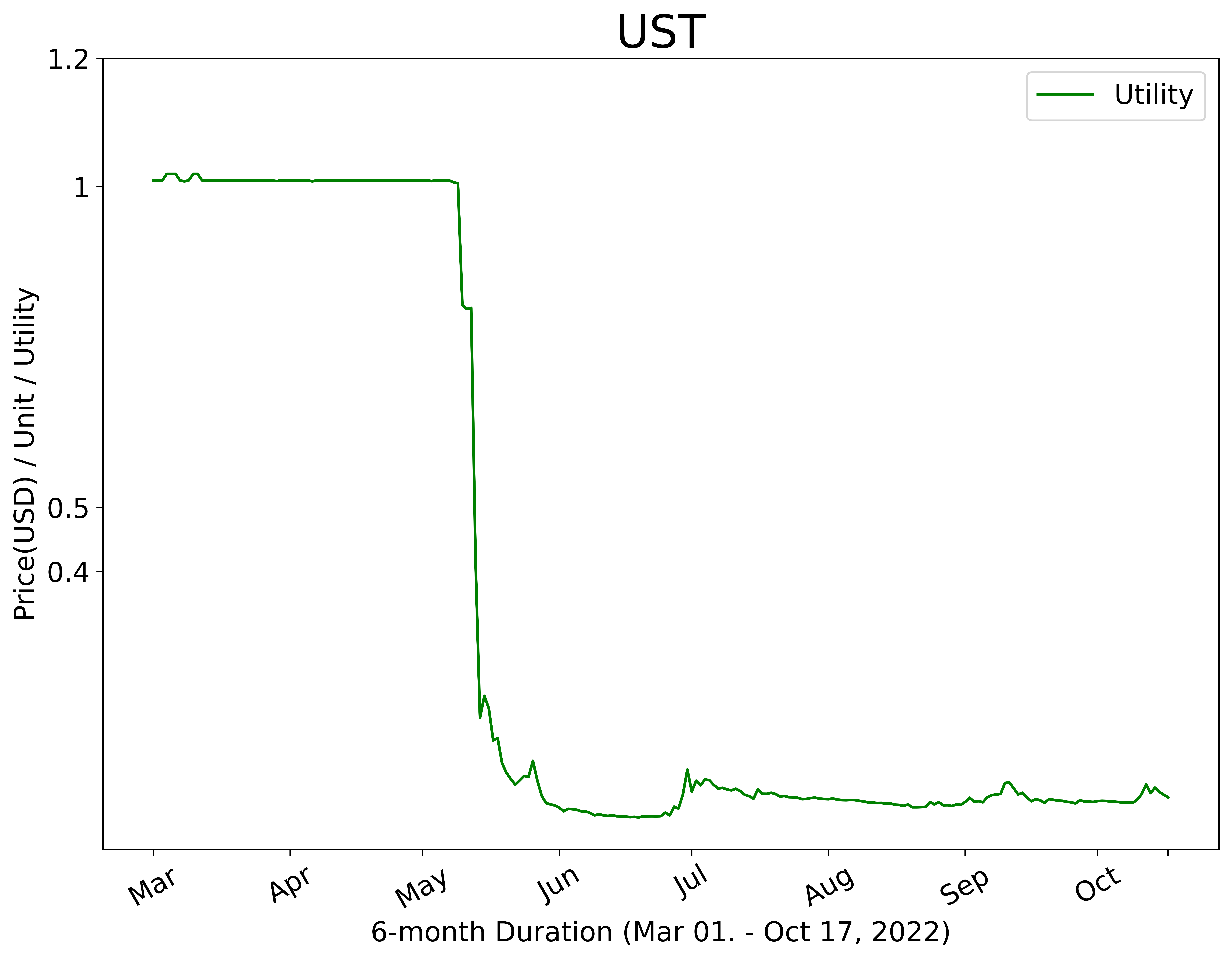}
	}
	}
	\caption{UST status over time.}
	\label{fig:ust}

\subfigure[BAC price]{\label{fig:bac1}
\resizebox{0.33\textwidth}{!}{
	\includegraphics[width=1\textwidth]{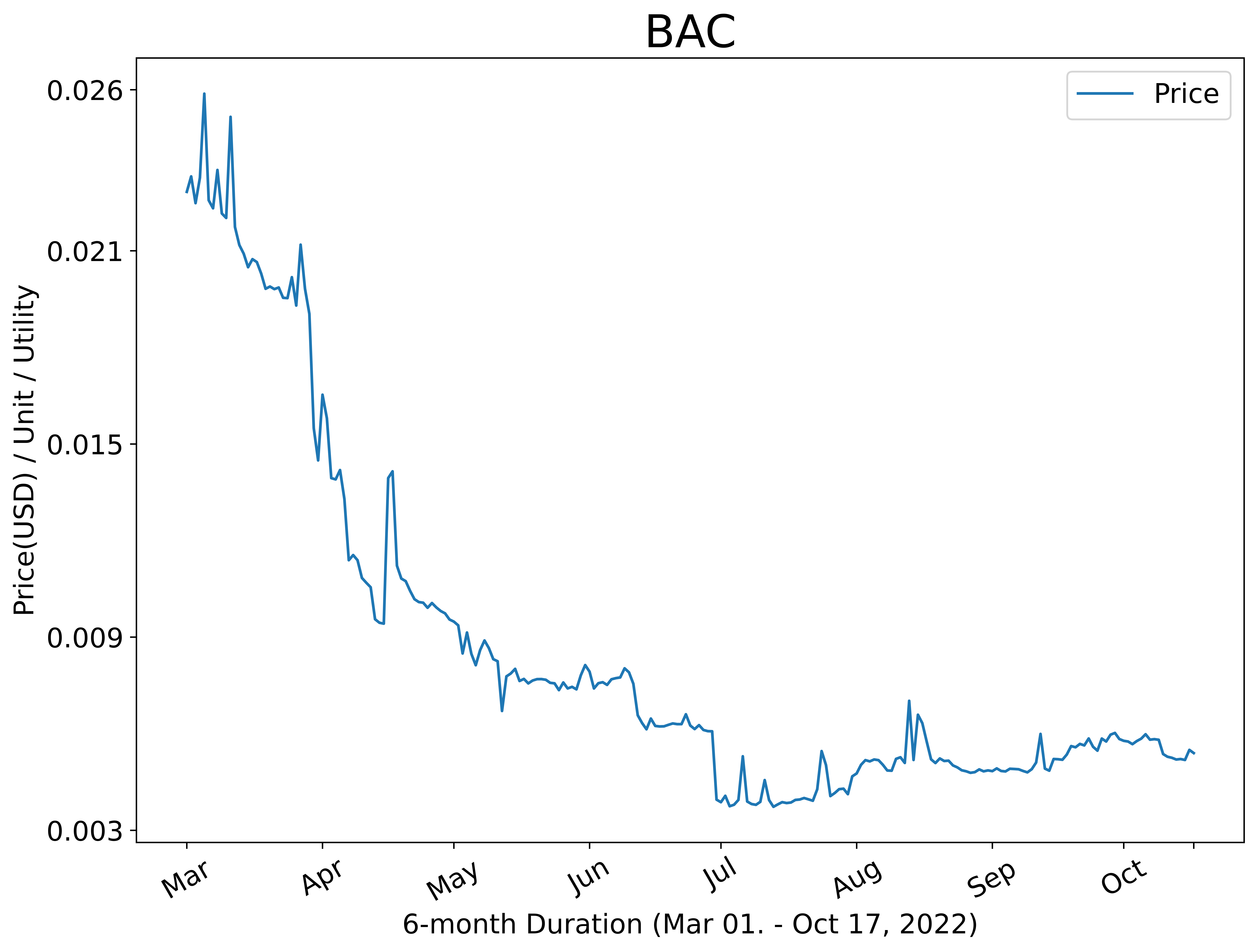}
	}
	}
\subfigure[BAC amounts]{\label{fig:bac2}
\resizebox{0.33\textwidth}{!}{
	\includegraphics[width=1\textwidth]{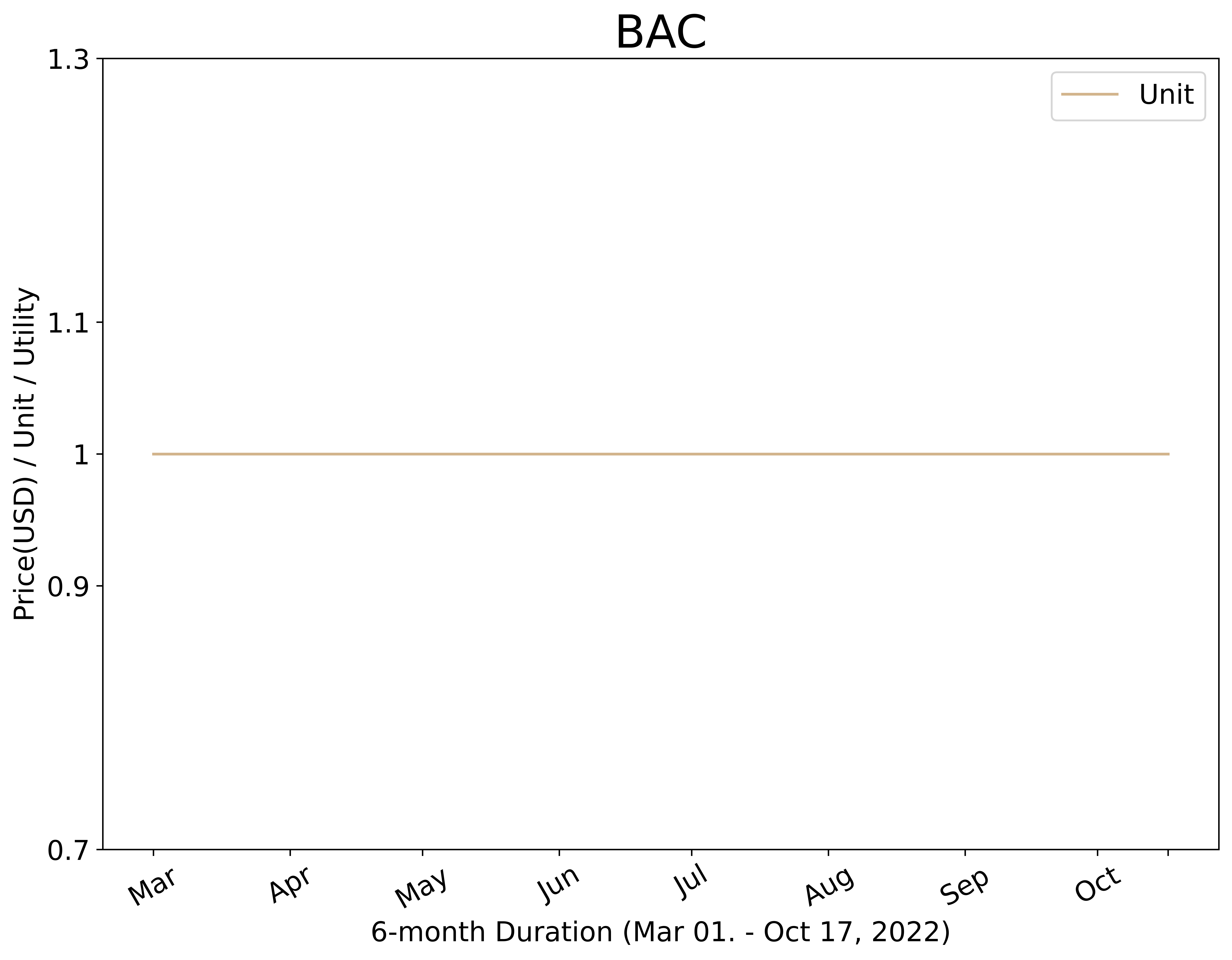}
	}
	}
\subfigure[Investor's utility]{\label{fig:bac3}
\resizebox{0.33\textwidth}{!}{
	\includegraphics[width=1\textwidth]{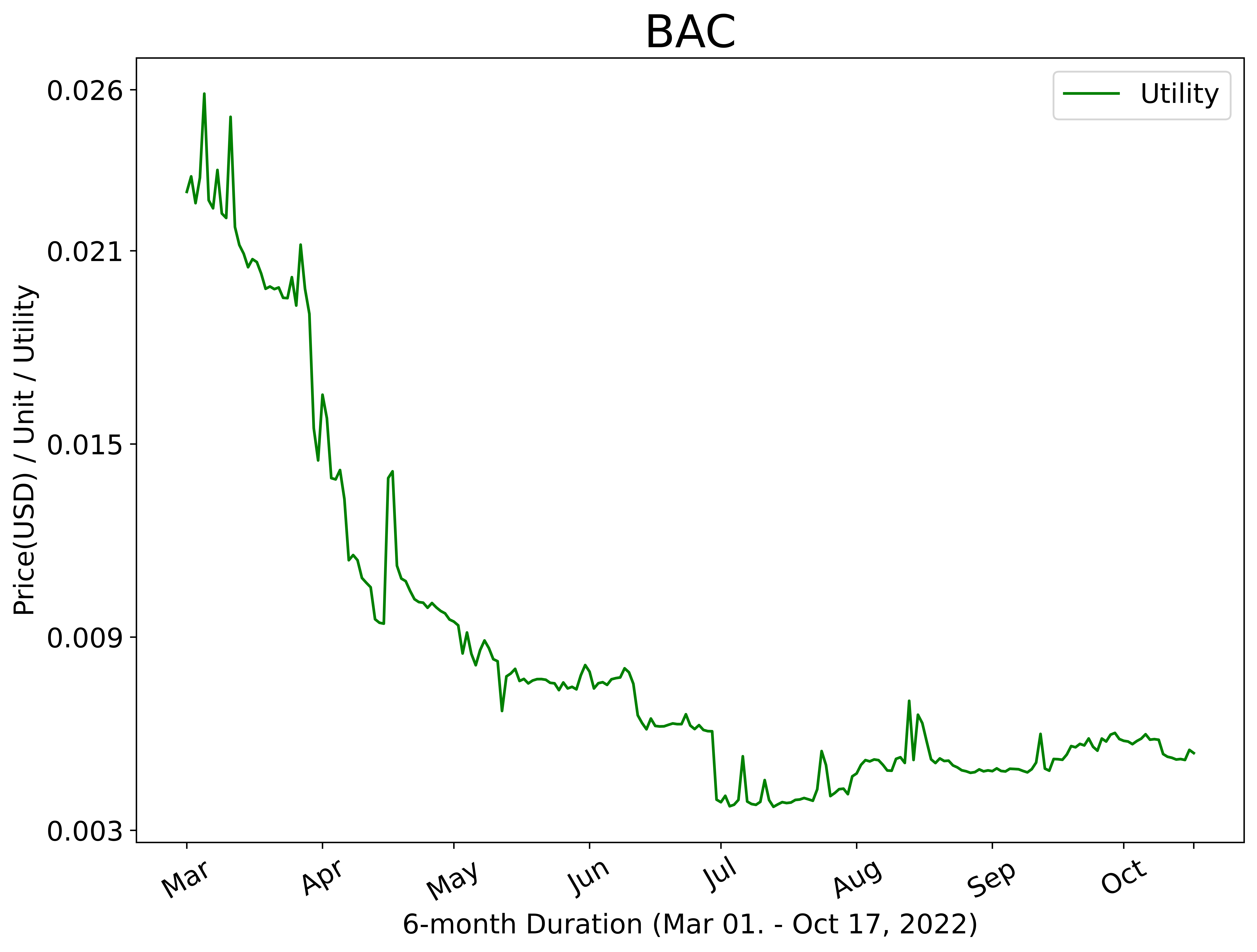}
	}
	}
	\caption{BAC status over time.}
	\label{fig:bac}
\end{figure*}

\smallskip
\noindent\underline{\textit{Investor's utility analysis.}}
Considering an investor who participates in the Ampleforth game at any time by purchasing one unit of AMPL, we analyze the utility changes over time. At each rebase time (cf. Fig.\ref{fig:ampl}), the investor's utility (green line) stays the same\footnote{Except for the rebase time in a day, the number of AMPL in wallets will not change. If the coin price rises and the investor selects to sell, his or her utility can be even greater than his original investment.}, meaning that the investor can always redeem (trade) the identical value of his investment back. However, the price of AMPL (blue line) is unstable, and the AMPL balance (brown line)\footnote{To better demonstrate the nature of rebase, we calculate daily AMPL numbers applying a simplest setting based on its 1st version whitepaper~\cite{amplwhiterpaper}.} is in the investor's wallet. In terms of a rational Ponzi model, AMPL has the potential to satisfy all listed conditions as long as there exists an lively traded market. 

Since any investor can retain at least the same utility when participating in the AMPL game, regardless of the token price, one may wonder why the rebase is not being adopted on a larger scale. The most obvious reason is due to its highly volatile price, resulting in holding AMPL being no different than holding a regular cryptocurrency. At the same time, one's balance varies proportionally to the price every day, making the holder fall into passive and anxious emotions. Besides, AMPL, as an ERC-20 token that works natively on Ethereum blockchain, cannot perform its rebase implementation smoothly in a centralized exchange or other blockchain platforms, significantly restricting its use scenarios. Ampleforth mitigates this constraint with the release of Wrapped-AMPL, a token that wraps AMPL similar to wrapped ETH. Wrapped-AMPL facilitates ecosystem integrations on both centralized and decentralized platforms.

\smallskip
\noindent\textbf{TerraUSD (Seiniorage Shares).}
Seniorage Shares is a monetary system design that is well-established in traditional finance.
The term ``seigniorage'' refers to the profits that are generated by a central bank through the creation of new money.
In the case of ``seigniorage'' shares stablecoins, this profit is generated by an algorithm that controls the supply of the stablecoin.
The seigniorage shares stablecoin model is designed to be self-regulating and can utilize a dual coin system to stabilize the value of one token.
The two coins are known as the ``shares'' and the ``price-stable coin''.
The share is designed to have a fluctuating value, much like a typical cryptocurrency.
Its value is determined by market demand and supply.
The price-stable coin, on the other hand, is pegged to a specific value.
The price-stable coin is used to stabilize the value of the shares, and it can be redeemed for the underlying asset at any time.
The two coins work together to provide a stable and flexible monetary system.
When the value of the shares starts to rise too fast, the price-stable coin is sold on the market, creating more supply and reducing the value of the shares.
When the value of the shares starts to fall too fast, the price-stable coin is redeemed for the underlying asset, reducing the supply and increasing the value of the share.
This dual coin design is intended to provide the benefits of a cryptocurrency while also providing the stability of a traditional currency.
It allows for market-based pricing of the shares while ensuring that the overall value of the currency remains relatively stable.
TerraUSD (UST) \cite{TerraUSD} is a seiniorage shares style stablecoin pegging to the USD on top of Terra's blockchain.
LUNA is the native staking token of the Terra blockchain, and it absorbs the volatility of UST as the token of shares.
Aibitrage can help making UST price stable: when the UST's price is $<1$ USD, arbitragers can send 1 UST to the system and receive 1 USD's worth of LUNA; whereas the UST's price becomes $>1$ USD, arbitragers will send 1 USD's worth of LUNA to the system and receive 1 UST.

\smallskip
\noindent\underline{\textit{Investor's utility analysis.}}
We also consider an investor who participates in the UST game at any time by purchasing one unit of UST. For simplicity, we assume that the investor will always hold UST without converting it to LUNA in the mid-run before exiting the game.  A chart of UST price (blue line), the number of UST in the wallet (brown line), and the investor's utility (green line) over time are presented in Fig.\ref{fig:ust}.
The practice proves the ineffectiveness of UST's stabilization as the UST eventually collapsed.
Investors suffered a significant loss from their initial investments, making UST unlikely to become a rational Ponzi game.

As a matter of fact, current dual-coin designs in the scope of seigniorage shares are practically difficult in realizing a rational model. These stablecoins are not substantially secured by anything, other than the expectation that the price-stable coin will eventually appreciate in value. Hence, a necessary condition for the seigniorage shares to work healthy is that the shares, or the secondary coin, must have some values. LUNA, as the secondary coin, should have, at least, trading values in this case, as people do use Terra blockchain every day. However, in extreme scenarios, investors may still lose confidence in it, and the massive sell-off has accelerated both two types of coins into a death spiral faster.

\smallskip
\noindent\textbf{Basis Cash (Seiniorage Shares).}
Seiniorage Shares systems in algorithmic stablecoin are not necessarily limited to dual-coin designs.
Basis Cash~\cite{basiscash} is also a seigniorage shares style protocol but with three tokens in the system, namely Basis Cash (BAC), Basis Bonds (BAB), and Basis Shares (BAS).
The protocol is designed to expand and contract supply similar to the way central banks trade fiscal debt to stabilize purchasing power.
Basis Cash (BAC), in this case, serves as the price-stable coin that maintains its peg to the MakerDAO Multi-Collateral Dai token.
In the previous examples, issuers either use one coin such as AMPL to rebase the user accounts, or apply a secondary token such as LUNA to absorb the volatility of the price-stable token of UST.
In the Basis Cash system, Basis Bonds (BAB) and Basis Shares (BAS) are both non-stable tokens that are designed to help stabilize the value of Basis Cash (BAC) by controlling its supply.
BAB are debt securities that can be minted and redeemed to incentivize changes in the BAC supply.
Bonds are always on sale to BAC holders.
The price of a Basis Bond is determined by market forces and can fluctuate based on demand.
When the price of BAC is below its peg, BAB can be purchased for less than the face value.
The bondholders can then redeem the bonds for the face value at a later date, earning a profit.
BAS are share tokens that loosely represent the value of the Basis Cash network.
When the price of BAC is above its peg, new Basis Shares are minted and sold to users, increasing the supply of BAS. Conversely, when the price of BAC is below its target price, existing BAS can be burned, reducing the supply of BAS.
This mechanism helps to stabilize the price of BAC over time.

\smallskip
\noindent\underline{\textit{Investor's utility analysis.}}
Due to the three-token design in Basis Cash, the complexity of the investor's possible operations increases dramatically.
For simplicity, we also apply the same setup as the previous examples. An investor can participate in the Basis cash game at any time by purchasing one unit of BAC. We assume that the investor will always hold BAC without converting it to BAB, and will not consider purchasing BAS in the mid-run before exiting the game.
Accordingly, a chart of price (blue line), the number of BAC in the wallet (brown line), and the investor's utility (green line) over time are presented in Fig.\ref{fig:bac}.
In practice, BAC, BAB, and BAS have all failed to maintain their coin prices and eventually went to almost zero in value.
Similar to the collapse of UST, BAC, as a more sophisticated seigniorage shares design, did not achieve a Rational Ponzi model result.

One thing worth noting is that BAC and BAS tokens are distributed through yield farming~\cite{basiscash}.
Yield farming, also referred to as liquidity mining, is a way to generate additional rewards by depositing your cryptocurrency into a pool.
In Basis cash, providing liquidity to BAC-DAI and BAS-DAI pairs result in additional BAS tokens being distributed.
This allows the participants in the game, to some extent, to gain tokens (utility) at no cost, and also can act as an incentive mechanism against lack of liquidity and bond death spirals.
In general, seigniorage shares designs are without any rebase or collateral risk, and there are theoretically infinite ways to split the volatility of the price-stable coin across multiple tokens.
This provides many possibilities to the algorithmic stablecoin practitioners with different expectations and risk preferences to design their customized monetary system.
However, seigniorage shares is inherently difficult to maintain the price stability, as the cases of UST and BAC have already demonstrated.


\section{Discussion}
\label{sec:discuss}

We now provide several additional thinking under the scope of this study in this section.

\smallskip
\noindent\textbf{Market cap vs price stability.}
Investors involved in stablecoin markets may often have an intuitive feeling that the larger the market cap of the stablecoin, the better stability the stablecoin can achieve with regard to its peg. However, according to the price and market cap data from CoinGecko \cite{coingecko}, this conjecture may not be valid as imagined for the case algorithmic stablecoins. The statement of ineffective application to the rebase design is explicit, as its market capitalization will vary as the supply of the currency is adjusted. For the seigniorage share design, this statement will be invalid as well. Due to its relative complexity, there are a lot of possibilities resulting in this situation, one of which is the loss of confidence. The recent joint collapse of LUNA/UST was a loop in which the market lost confidence and entered a death spiral, pushing them out of the market cap top-10 within merely several hours. Thus, a larger market cap of an algorithmic stablecoin does not necessarily guarantee a better stability.

\smallskip
\noindent\textbf{Price oracle risk.}
Price oracle provides the price feed for on-chain DeFi protocols \cite{werner2021sok}, and it is crucial to the secure operation of an algorithmic stablecoin. Algorithmic stablecoins may have special price-feeding needs according to their different algorithms, which may require the establishment of customized oracles.
Constructing an internal oracle is one way, but this may lead to an additional layer with potential centralization and incentivize the stakeholder to update the price in his favor.
Opting for an external price oracle can mitigate this problem, but it is still vulnerable to external price attacks include flash loan price manipulation attack, pump, and dump~\cite{Attackflashloan}.
Selecting a trustworthy price oracle is not easy for an algorithmic stablecoin, and a comprehensive ex-ante risk assessment is required.

\smallskip
\noindent\textbf{Reconsider the necessity of a peg.}
Stablecoins are designed to be pegged to some underlying assets, with the USD being a popular choice. Anchoring a benchmark is an effective way to achieve stability in stablecoins. People may thus generally assume a fact that the stablecoin is perfectly fungible for its peg. However, this is not the full truth. Stablecoins can be, in fact, rather considered as their own free-floating assets that closely track the value of a peg. Besides, it is also possible that stablecoins could provide the desired stability without pegging to an asset. In one possible future, once the economy develops surrounding the stablecoin itself, and such stablecoins have also been widely used as a medium of exchange, maintaining a perfect peg becomes unnecessary.

\smallskip
\noindent\textbf{Rational Ponzi in DeFi.}
At a high level, DeFi naturally has an advantage in realizing a rational Ponzi game.
In traditional finance, default has always been a tough problem to deal with.
Default is a situation in which a borrower is unable to meet their financial obligations to a lender, such as failing to make scheduled loan payments.
In other words, default occurs when a borrower cannot pay back their debt as agreed.
However, DeFi protocols are built on top of blockchain technology and are designed to be trustless, meaning they do not rely on intermediaries such as banks or other financial institutions to manage transactions.
Instead, of leveraging automated smart contracts, DeFi has many advanced and innovative financial tools such as automated market makers (AMM), flash loans, and enforceable payment.
This ensures all parties involved in a game adhere to the terms of the agreement and thus provides more possibility to run a rational Ponzi in DeFi successfully.

\section{Conclusion}
\label{sec:conclu}

This study applies O'Connell \& Zeldes's model (1988, cf.\cite{ponzi_rational}) for rational Ponzi games to the context of algorithmic stablecoins.
Following its connotation, we establish a rational Ponzi model as the measurement and accordingly evaluate three mainstream algorithmic stablecoin projects (Ampleforth, TerraUSD, and Basis Cash) as instances.
Our investor utility analysis indicates that the rebase design has the potential to realize a rational Ponzi game.
However, the seigniorage share approach seems hard to do so.
It is seldom possible to run a rational Ponzi in seigniorage share styled algorithmic stablecoins, as once entering a death spiral, the selling pressure in the case of a multi-coin design is intrinsically uncontrollable.
But in general, implementing rational models in DeFi is always easier than in traditional finance.
Beyond that, we also point out the hurdles existing in current algorithmic stablecoin implementations and identify directions for future endeavors.

{\footnotesize \bibliographystyle{IEEEtran}
\bibliography{bib}}

\end{document}